\documentclass[pdflatex,sn-mathphys-num]{sn-jnl}

\usepackage{graphicx}%
\usepackage{multirow}%
\usepackage{amsmath,amssymb,amsfonts}%
\usepackage{amsthm}%
\usepackage{mathrsfs}%
\usepackage[title]{appendix}%
\usepackage{xcolor}%
\usepackage{textcomp}%
\usepackage{manyfoot}%
\usepackage{booktabs}%
\usepackage{algorithm}%
\usepackage{algorithmicx}%
\usepackage{algpseudocode}%
\usepackage{listings}%

\theoremstyle{thmstyleone}%
\theoremstyle{thmstyletwo}%

\theoremstyle{thmstylethree}%

\begin{document}

\title[NMR on Ba$_2$La$_2$CoTe$_2$O$_{12}$]{NMR study on equilateral triangular lattice antiferromagnet Ba$_2$La$_2$CoTe$_2$O$_{12}$}


\author[1]{\fnm{Keito} \sur{Morioka}}

\author*[1]{\fnm{Takayuki} \sur{Goto}}\email{gotoo-t@sophia.ac.jp}

\author[2]{\fnm{Masari} \sur{Watanabe}}

\author[2]{\fnm{Yuki} \sur{Kojima}}

\author[2]{\fnm{Nobuyuki} \sur{Kurita}}

\author[2]{\fnm{Hidekazu} \sur{Tanaka}}

\author[3]{\fnm{Satoshi} \sur{Iguchi}}

\author[3]{\fnm{Takahiko} \sur{Sasaki}}

\affil*[1]{\orgdiv{Physics Division}, \orgname{Sophia University}, \orgaddress{\street{Kioi-cho 7-1}, \city{Chiyoda-ku}, \postcode{1028554}, \state{Tokyo}, \country{Japan}}}

\affil[2]{\orgdiv{Department of Physics}, \orgname{Institute of Science Tokyo}, \orgaddress{\street{Oh-okayama}, \city{Meguro-ku}, \postcode{152-8551}, \state{Tokyo}, \country{Japan}}}

\affil[3]{\orgdiv{Institute for Materials Research}, \orgname{Tohoku University}, \orgaddress{\street{Katahira}, \city{Aoba-ku}, \postcode{980-8577}, \state{Sendai}, \country{Japan}}}


\abstract{
\unboldmath
We report a $^{139}$La-NMR study of Ba$_2$La$_2$CoTe$_2$O$_{12}$, $S=1/2$ equilateral triangular-lattice 
antiferromagnet with easy-plane anisotropy at low temperatures.  
This compound undergoes a magnetic phase transition at $T_{\rm N} =$ 3.26 K into an ordered state with the {120}$^\circ$ spin structure.  
Under magnetic fields above 3T, $T_{\rm N}$ splits into $T_{\rm N1}$ and $T_{\rm N2}$, which correspond 
to the transitions from the paramagnetic phase to the up-up-down (uud) phase and from the uud phase to the triangular coplanar phase, respectively. 
The NMR spin-lattice relaxation rate $1/T_1$ exhibits a critical divergence at $T_{\rm N1}$, indicating the onset of long-range magnetic order. 
At $T_{\rm N2}$, the NMR-linewidth measured at 5.4 T exhibits an anomalous decrease, which 
we attribute to a change in the spin structure from the uud to the triangular coplanar phase.

}


\maketitle

\section{Introduction}\label{sec1}

In quasi-two-dimensional spin systems, geometrical frustration and quantum fluctuation often give 
rise to novel and exotic spin states at low temperatures.
Among them, the one-third magnetization plateau in the triangular-lattice Heisenberg antiferromagnet (TLHAF) 
\cite{Collins,Kamiya_neutron,Nakajima_Goto}
attracts considerable interest 
as a manifestation of quantum effects \cite{BCSO_Shirata,Nikuni_Shiba,Kawamura_Miyashita}.
Ba$_2$La$_2$CoTe$_2$O$_{12}$ is an $S =$1/2 quasi-two-dimensional antiferromagnet on the equilateral triangular lattice, 
and has recently been reported 
to exhibit a 1/3-magnetization plateau in the field region between 8.7 and 15.2 T at 1.3 K, 
where the 
system is expected to be in the collinear up-up-down (uud) 
phase \cite{BLCTO_Kojima,Field-orientation-dep,ultrasound}.

Previous studies have reported  
that the nearest-neighbor exchange is $J =$ 22 K 
and that the interlayer spacing is 9.2 \AA, large enough to be considered as a good quasi two  
dimensional system. 
Existence of the substantial planar XXZ anisotropy has been reported by magnetic susceptibility and 
inelastic neutron scattering \cite{NatureComm_anisotropic,BLCTO_Kojima}.
The uniform susceptibility increases with decreasing temperature below 50 K and exhibits a rounded maximum at 5 K, which
is characteristic of a low-dimensional antiferromagnet.
In zero field, the system shows a long-range magnetic order with the 120$^\circ$ structure
at $T_{\rm N} =$ 3.26 K.
By the neutron-diffraction measurements, the magnitude of the ordered moment is 
reported to be 0.63 $\mu_{\rm B}$ at 1.6 K \cite{BLCTO_Kojima}.
Under magnetic fields above 3 T, the successive phase transitions were observed at
$T_{\rm N1}$ and $T_{\rm N2} (< T_{\rm N1})$ as two distinct peaks of the specific heat \cite{BLCTO_Kojima}. 
They attributed this second peak at $T_{\rm N2}$ to the possible spin structure transition from the uud to
the triangular coplanar phase \cite{BLCTO_Kojima,Kawamura_Miyashita,Nikuni_Shiba}.

The purpose of this work is to investigate the spin state within and outside of the plateau by the microscopic probe of NMR. In the following, we investigate the $^{139}$La-NMR spectra and the nuclear spin-lattice relaxation rate $1/T_1$ to show that Ba$_2$La$_2$CoTe$_2$O$_{12}$
undergoes successive phase transitions from the paramagnetic state through the uud and finally 
to the triangular coplanar phase under the limited field region.

\section{Experimental}\label{sec2}

\begin{figure}[h]
\centering
\includegraphics[width=0.8\textwidth,bb=25 410 580 800,clip]{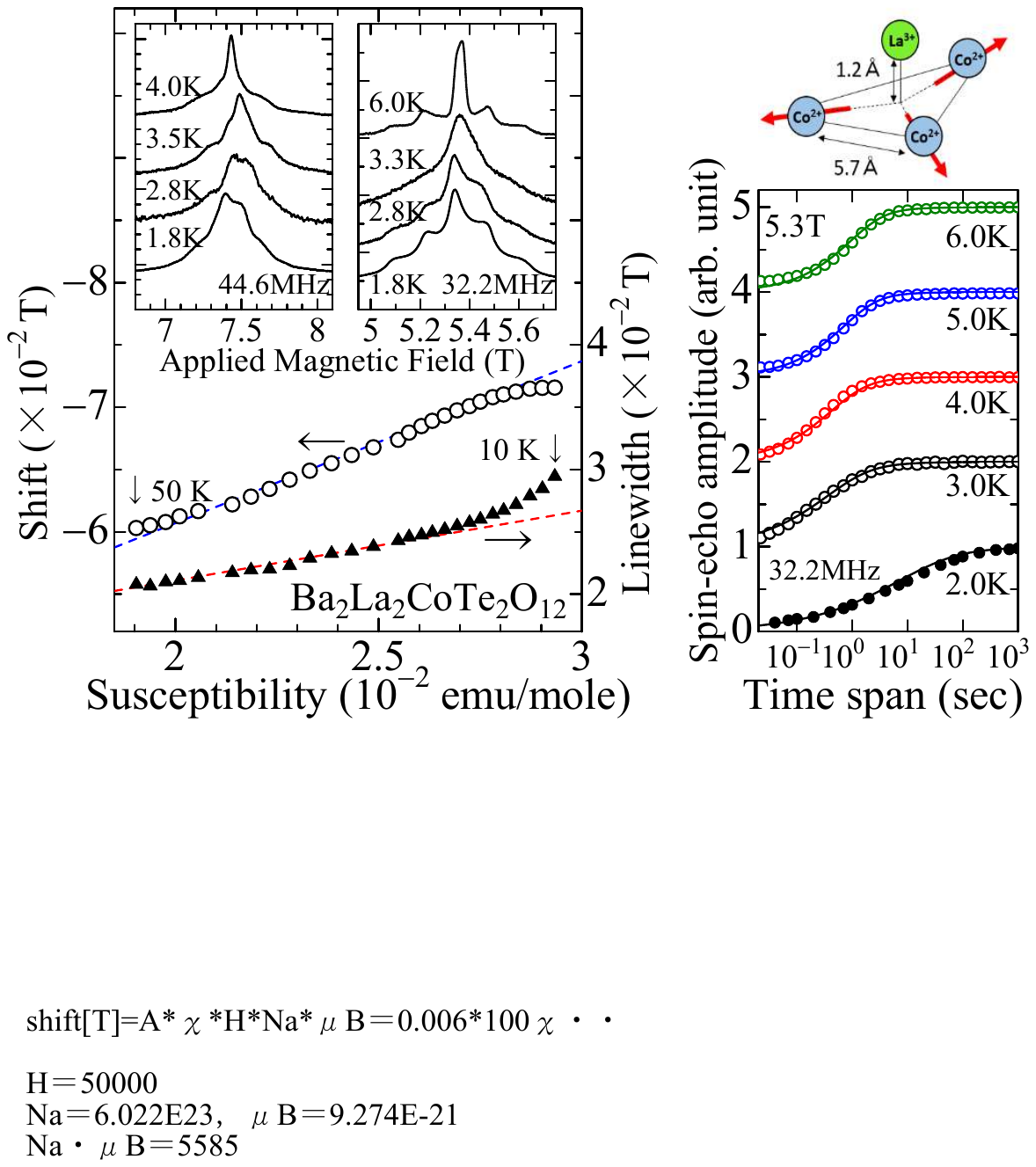}

\bigskip
\caption{
(Left) NMR shift and linewidth versus magnetic susceptibility in
the paramagnetic region 10 - 50 K.
The dashed lines show the isotropic and anisotropic part of hyperfine coupling constans
$A_{\rm iso} = -660$ and $3A_{\rm an} = +190$ (Oe/$\mu_{\rm B}$), respectively.
Insets show typical spectra at low temperatures.
The $eqq$ interaction parameter was obtained from the distances between the singular points,
to be $^{139}\nu_{\rm Q} =$ 1.43 MHz.
(Top Right) Schematic of the local structure around the NMR site La.
(Bottom Right) Spin-echo amplitude versus time span between the saturating pulse train and 
the echo-measurement pulses. Data for each temperature is shifted vertically.  
Solid curves show the theoretical recovery function for $I = 7/2$ nuclei shown in the text.
}\label{fig1}
\end{figure}

$^{139}$La-NMR was measured on a powder sample of Ba$_2$La$_2$CoTe$_2$O$_{12}$ \cite{BLCTO_Kojima} 
at low temperatures down to 1.8 K and under the fields between 4 and 9 T.  
Spectra were obtained by plotting the spin-echo amplitude against the applied field, 
which was slowly changed within the field region of spectrum.
As shown in the inset of Fig. \ref{fig1}, 
the La site is located $1.2$ \AA \hspace{1mm} above the center 
of an equilateral triangle formed by three Co$^{2+}$ ions.
Hence, $^{139}$La-NMR is expected to probe the local magnetization averaged over the triangle, through
both the hyperfine interaction and the classical dipole-dipole interaction.

The inset in Fig. 1 shows the typical $^{139}$La-NMR spectra at various temperatures.
In the paramagnetic state of 6 K, singular points in the powder pattern due to the nuclear quadrupole 
interaction ($eqq$) are clearly seen 
for all the nuclear transitions of $^{139}$La($I=7/2$), 
assuring the sample quality. 
The linewidth was determined from the central or $I_z = \pm1/2$ transition peak 
with the definition of fullwidth at 70 \% maximum intensity.
The nuclear spin-lattice relaxation rate $1/T_1$ was measured by the saturation-recovery method for the central peak.
The recovery curves were analyzed by the formula 
$1-\frac{1}{84}e^{-(t/T_1)^\beta}-\frac{3}{44}e^{-6(t/T_1)^\beta}-\frac{75}{364}e^{-15(t/T_1)^\beta}-\frac{1225}{1716}e^{-28(t/T_1)^\beta}$, 
where $t$ is the distance between the pulse train for saturation and echo-generating $\pi/2-\pi$ pulses 
and $\beta$ is the stretched exponent
reflecting the distribution of relaxation times due to 
sample inhomogeneity, which was set to be approximately 
0.4 \cite{Stretched_DC_Johnston,La-recovery-Furukawa,Narath,Watanabe_Goto}.
Typical recovery curves are shown in Fig. 1.
Here, the value of $\beta$ suggests considerable inhomogeneity in the sample. 
In this type of oxide, inter-substitution between La and Ba ion may be possible. 
However, its amount is considered to be small, because eqq-singular points are clearly observed in 
NMR spectra in the paramagnetic state, and no sprious-Curie term is observed in the
uniform susceptibility \cite{BLCTO_Kojima}.


\begin{figure}[h]
\centering

\includegraphics[width=0.8\textwidth,bb=10 390 585 790,clip]{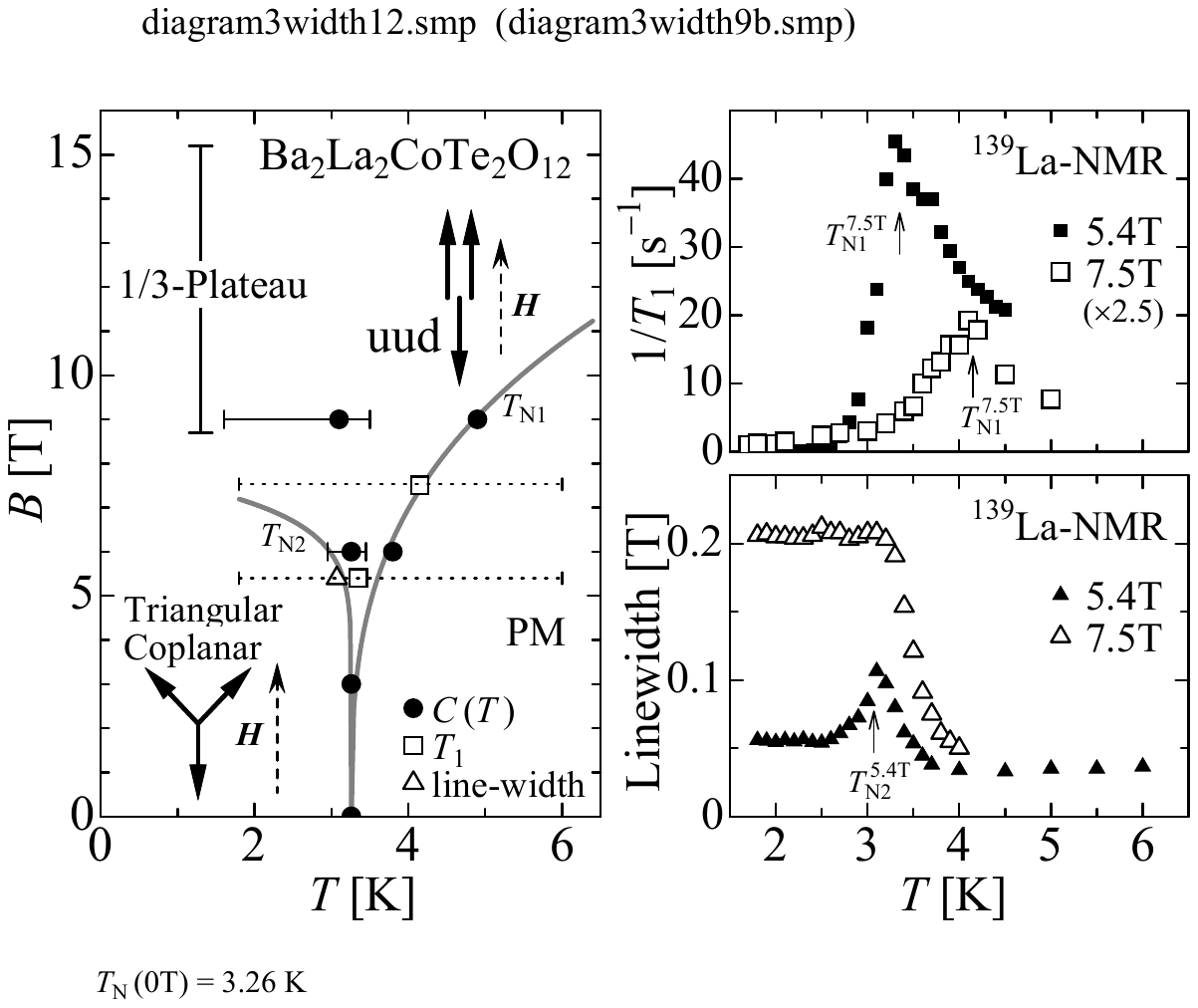}
\bigskip
\caption{(left) B-T phase diagram for the magnetic field parallel with the triangular plane, 
expected from specific heat and magnetization measurements.  
The error bars of specific heat come from the fact that observed peaks at high fields are broad \cite{BLCTO_Kojima}.
The position of the one-third plateau experimentally determined is shown by the vertical 
bar \cite{BLCTO_Kojima}.  
Gray solid curves show the expected phase boundaries between the paramagnetic phase and uud ($T_{\rm N1}$) and
between the uud and the triangular coplanar phase $T_{\rm N2}$. 
Thick solid and dashed arrows show the schematic spin structures and the dirrections of applied fields for each phases. 
Horizontal dotted arrows show the experimental conditions of temperature range and of magnetic fields.
(right) Temperature dependence of $^{139}$La-NMR $1/T_1$ and linewidth measured under the magnetic fields at 
around 5.4 and 7.5 T.
Vertical arrows show maxima of each quantities.
}\label{fig2}
\end{figure}

\section{Results and Discussion}\label{sec3}

Figure 2 (right) shows the temperature dependence of $1/T_1$ and the spectral linewidth measured under 
applied fields of approximately 5.4 and 7.5 T.  
As the temperature is lowered,  $1/T_1$ exihibts a critical divergence 
at the phase transition from the paramagnetic to the magnetically ordered state.
Observed fields and temperatures agreed with the results of specific heat \cite{BLCTO_Kojima} 
as seen in Fig. \ref{fig2} (left),
assuring that the system shows the long-range magnetic order at $T_{\rm N1}$.
NMR signal intensity showed a slight reduction in the vicinity of $T_{\rm N1}$, suggesting
that $T_2$, the transverse nuclear spin relaxation time is decreased due to the
critical slowing down.

Next, the temperature dependence of linewidth exhibited quite different behavior
depending on the applied field.  
With decreasing temperature, the width starts to increase nearly at $T_{\rm N1}$.  
At 7.5 T, it stays constant down to 1.8 K, keeping the large value, a behavior commonly 
observed in typical antiferromagnets \cite{Watanabe_Goto}.
The increment from the paramagnetic state to the lowest temperature is approximately 1500 Oe,
which is consistent with the ordered moment of 0.63 $\mu_{\rm B}$ \cite{BLCTO_Kojima} and
the obtained hyperfine coupling constant -660 Oe, if one recalls the fact 
that the magnitude of ordered moment obtained by the neutron experiment gives 
in generall a lower limit value.

At 5.4 T, temperature dependence of the width is quite different.
It shows a peak at $T_{\rm N2}$ and decreases again in the lower temperatures.
Since the magnitude of the ordered moment at each Co site increased monotonically upon 
cooling in the ordered state as was observed by neutron diffraction experiments \cite{BLCTO_Kojima}, 
the observed decrease in the linewidth must originate from a 
change in the spin structure rather than a reduction of the ordered moment.
We consider this reduction at 5.4 T reflects the change in spin structure, 
from the uud to the triangular coplanar phase as pointed out by Kojima \cite{BLCTO_Kojima}.
On the other hand, in the measurement at 7.5 T, the linewidth stayed constant,
indicating that the system stays in uud phase, at least down to 1.8 K.
We also mention that preliminary measurements at 15 T, which is well inside the 
plateau region, showed no reduction in the linewidth at low temperatures.
This also assures that the linewidth reduction is observed only when the system
gets into the triangular coplanar phase from uud.

In order to prove the detailed relation between the linewidth and spin structures,
a numerical simulation for the NMR powder pattern is necessary, which is
now under progress with the measurements on aligned powder samples.
At qualitative level, one can understand this reduction in terms of the geometrical 
cancellation of ordered moments within a triangle.  
That is, net field from the three spins surrounding La-site 
in the coplanar phase is apparently smaller than that in the uud phase.  
Note that for the quantitative discussion, one must take into account the
interaction between ordered moments and the La nucleus, and also the
effect of applied field to the spin direction.

In contrast to the linewidth, $1/T_1$ exhiibits a single peak at  $T_{\rm N1}$ and 
decreases monotonically 
down to the lowest temperature, irrespective of the applied field of 5.4 or 7.5 T.
In particular, it was not affected by the second transition $T_{\rm N2} =$ 3.1 K  at 5.4 T.
This is because the transition at $T_{\rm N2}$ occurs between the two statically ordered phase, 
and therefore not accompanied by critical slowing down.

We believe that the use of a powder sample does not affect our main conclusions. 
In a randomly oriented powder, crystallites with the triangular plane nearly parallel to 
the applied field—which host the 1/3-magnetization plateau—dominate over those 
with perpendicular orientations in a statistical ratio of approximately 1:2. 
This is consistent with the clear observation of successive phase transitions and the 1/3 plateau 
in powder measurements \cite{BLCTO_Kojima}..

Finally, in order to complete the phase diagram and to explain quantitatively the reduction in the observed linewidth 
below $T_{\rm N2}$, NMR measurements in wider field region is necessary. 
The behavior of $1/T_1$ at still lower temperature is expected to give an important information on 
the spin excitation gap or the magnon properties. We are now planning these measurements for comprehensive
understanding of this system.

\section{Conclusion}

NMR measurements were performed on the equilateral triangular antiferromagnet Ba$_2$La$_2$CoTe$_2$O$_{12}$.
$1/T_1$ showed critical divergence at $T_{\rm N1}$, demonstrating a second-order phase transition to
the magnetically ordered state. 
At 5.4 T, the linewidth increased below $T_{\rm N1}$,
but decreased again below $T_{\rm N2}$,
indicating that the system
undergoes successive phase transitions culminating in the triangular coplanar phase.

\bmhead{Acknowledgements}

This work was supported by JSPS KAKENHI Grant Number 25H01546 and 25K07210.
A part of this work was performed at the Institute for Materials Research, Tohoku University (Project No 202412-RDKGE-0046 and 202412-HMKPB-0037).  





\bibliography{lt30_B2L2CT2O12_NMR_251212.bib}

\end{document}